\documentstyle[twocolumn,eqsecnum,aps]{revtex}

\begin{document}
\draft
\title{Efficient schemes for reducing imperfect
collective decoherences
 }
\author{WonYoung Hwang \cite{email}, Hyukjae Lee,
 Doyeol (David) Ahn \cite{byline},
 and Sung Woo Hwang \cite{byline2}}

\address{ Institute of Quantum Information Processing
and Systems,
 University of Seoul 90, Jeonnong, Tongdaemoon,
 Seoul 130-743, Korea
}
\maketitle
\begin{abstract}
We propose schemes that are efficient when each pair of
qubits undergoes some imperfect collective decoherence with
different baths. In the proposed scheme, each pair of
qubits is first encoded in a decoherence-free subspace
composed of two qubits. Leakage out of the encoding space
generated by the imperfection is reduced by the quantum Zeno
effect. Phase errors in the encoded bits generated by the
imperfection are reduced by concatenation of the
decoherence-free subspace with 
either a three-qubit quantum error
correcting code that corrects only phase errors or
a two-qubit
quantum error detecting code that detects only phase
errors, connected with the quantum Zeno effect again.
\end{abstract}
\pacs{03.67.Lx, 03.65.Bz}

\narrowtext

Information processing with quantum bits (qubits), e.g.,
quantum computing, quantum cryptography, and quantum
gambling is, a novel technique that solves some classically
intractable problems \cite{feyn}-\cite{gold}. However, in
order to make quantum information processing involving many
qubits practical, some methods for reducing
decoherence (MRDs) are indispensable. Among these, there are
quantum error correcting codes (QECCs)
\cite{shor}-\cite{pres}, decoherence-free subspaces (DFSs)
\cite{chua}-\cite{lida}, the quantum Zeno effect (QZE)
\cite{vaid,dua2}, \footnote{The 
QZE was discovered by Misra and
Sudersan \cite{misr}. The use of the QZE for combating
decoherence was first suggested by Zurek \cite{zure}, and it
is a part of a scheme considered by Barenco et al.
\cite{bare}.} and dynamical suppression of decoherence
\cite{viol}.

If the provisos for DFSs are fulfilled, DFSs are more
efficient than QECCs or the QZE in respect to the
amount of other
necessary resources as well as the number of qubits.
The robustness
of DFSs against perturbation of replica symmetry
is shown in Refs. \cite{baco,lida}. Indeed, qubits in
collective decoherence with imperfect replica symmetry can
be preserved with concatenation of DFSs with QECCs
\cite{lid2}.
 However, the efficiencies of various MRD's 
depend on the decoherence model. So devising an optimal
scheme that appropriately combines existing MRDs for a
given decoherence model will be important in the design of
quantum information processors. In this paper, we propose a
scheme that is efficient when each pair of qubits undergoes
imperfect collective decoherence with different baths
(cluster decoherence \cite{lida,lid2}). We start with a
subspace composed of two qubits which is decoherence-free
against a certain interaction that generates only phase
errors. Other interactions assumed small but non-negligible
make the encoded states leak out of the DFS, and the
interactions generate phase errors in the encoded qubits. The
leakage is reduced by the QZE. The phase errors in the encoded
qubits are corrected by concatenating the DFS 
with a three-qubit
QECC that corrects only phase errors \cite{knil} or by
concatenating the DFS with a two-qubit quantum error detecting
code that detects only phase errors and by QZE again.

The dynamics of the qubits and bath is governed by
\begin{eqnarray}
{\bf H}_T={\bf H}_S + {\bf H}_B +{\bf H}_I,
\end{eqnarray}
where ${\bf H}_T$, ${\bf H}_S$, and ${\bf H}_B$ denote the
total, the system, and the bath (or environment)
Hamiltonian, respectively,
and ${\bf H}_I$ is the interaction Hamiltonian. First, we
consider the following simple model.
\begin{eqnarray}
{\bf H}_S&=& \epsilon(\sigma_1^z +  \sigma_2^z),
\nonumber\\
{\bf H}_I&=& \lambda (\sigma^z_1 +\sigma^z_2)\otimes V_z,
\label{ab}
\end{eqnarray}
where $\sigma^z_i$ ($i=1,2$) are Pauli spin operators, 
$V_z$ is the bath operator coupled to the degree of freedom,
and ${\bf H}_B$ is arbitrary.
The type of Hamiltonian in Eq. (\ref{ab}),
which corresponds to a special case of the
spin-boson problem, has been used by many authors to model
decoherence despite its simplicity
\cite{viol,unru,palm}:
This model describes a decohering mechanism with only
phase errors. Amplitude
errors would involve a time scale much longer than that
of phase errors in some real physical systems
\cite{palm}.
(Later, we will
treat more general models.) We can easily see 
that a subspace spanned by 
$|01\rangle$ and $|10\rangle $ satisfies the DFS
condition in the case of the interaction given by
Eq. (\ref{ab}): $(\sigma^z_1 +\sigma^z_2)|01\rangle = 0
|01\rangle$ and $(\sigma^z_1 +\sigma^z_2)|10\rangle = 0
|10\rangle$.
\footnote{It is noted that  another condition should be
additionally satisfied in order that some subspaces become
decoherence-free: the system Hamiltonian ${\bf H}_S$ 
does not
make qubits leak out of the subspace. Otherwise, ${\bf
H}_S$ needs to be eliminated to satisfy this condition by the
method proposed in Ref. \cite{duan}.} Therefore,
\begin{eqnarray}
{\bf H}_T [(\alpha |01\rangle + \beta|10\rangle)
\otimes && |\Psi_{b}(0)\rangle ] \nonumber\\
= (\alpha |01\rangle + \beta|10\rangle) \otimes &&
[0(\epsilon + \lambda V_z)+{\bf H}_B]
|\Psi_{b}(0)\rangle,
\label{ba}
\end{eqnarray}
and as a result
\begin{eqnarray}
\exp[-i{\bf H}_T T_0]
(\alpha |01\rangle + \beta|10\rangle)
\otimes |\Psi_{b}(0)\rangle \nonumber\\
= (\alpha |01\rangle + \beta|10\rangle) \otimes
  \exp[-i{\bf H}_B T_0] |\Psi_{b}(0)\rangle.
\label{bb}
\end{eqnarray}
Here we can see that the qubits indeed do not decohere. So
Span$[|01\rangle,|10\rangle ]$ (the subspace that
$|01\rangle$ and $|10\rangle $ span) can be used to encode
one qubit. That is, we can encode a qubit $\alpha |0\rangle
+ \beta |1\rangle$ into, for example, the state
\begin{equation}
|\Psi_{enc}\rangle= \alpha |01\rangle + \beta |10\rangle.
\label{ad}
\end{equation}
It is clear that this subspace is sufficient for preventing
decoherence provided that the system and bath are perfectly
governed by Eq. (\ref{ab}). However, in real systems there
are some small perturbative interactions that are not
included in Eq. (\ref{ab}). When the perturbative interaction
is non-negligible, its effect must be reduced by some method
that we will describe. Now, let us consider a more general
decoherence model:
\begin{eqnarray}
{\bf H}_S&=& \epsilon_1 \sigma_1^z +  \epsilon_2 \sigma_2^z,
\hspace{1cm}
\nonumber\\
{\bf H}_I&=&
 [(\lambda_1^z \sigma_1^z + \lambda_2^z \sigma_2^z)
  \otimes V_z +
  (\lambda_1^+ \sigma_1^+ + \lambda_2^+ \sigma_2^+)
  \otimes V_+ +
\nonumber\\ &&
  (\lambda_1^- \sigma_1^- + \lambda_2^- \sigma_2^-)
  \otimes V_- ].
  \label{ae}
\end{eqnarray}
Here,  $\sigma_i^j$ ($j= z,+,-$) are Pauli spin
operators and $V_j$ are the bath operators coupled to these
degrees of freedom. We assume that $\Delta \epsilon \equiv
\epsilon_2-\epsilon_1\ll\epsilon_1$
and $\Delta \lambda^z \equiv
\lambda_2^z-\lambda_1^z \ll\lambda_1^z$.
We also assume that phase
damping is dominant $\lambda_i^z \gg \lambda_i^{+}$ and
$\lambda_i^z \gg \lambda_i^{-}$. In the limit when $\Delta
\epsilon$, $\Delta \lambda^z$,
$\lambda_i^{+}$, and $\lambda_i^{-}$ vanish, Eq. (\ref{ae})
reduces to Eq. (\ref{ab}).
Then let us consider the following. In a short period of
time $T_0/N$, under the Hamiltonian ${\bf H}_T$, the encoded
state evolves into
\begin{eqnarray}
&&|\Psi(T_0/N)\rangle
\nonumber\\
&\approx&
[1-i{\bf H}(T_0/N)](\alpha
|01\rangle + \beta|10\rangle)
\otimes |\Psi_{b}(0)\rangle \nonumber\\
&=& (\alpha |01\rangle + \beta
|10\rangle) \otimes
\nonumber\\ 
&&[1- 0 i(T_0/N)
(\epsilon_1 + \lambda_1^z V_z)
-i(T_0/N){\bf H}_B]|\Psi_{b}(0)\rangle
\nonumber\\
 && +(T_0/N)(-\alpha |01\rangle+\beta |10\rangle)\otimes
(\Delta \epsilon + \Delta \lambda^z V_z)|\Psi_{b}(0)\rangle
\nonumber\\ && +(T_0/N)|00\rangle \otimes (\lambda_1^+ \beta +
\lambda_2^+ \alpha )V_+ |\Psi_{b}(0)\rangle \nonumber\\&&
+(T_0/N)|11\rangle \otimes (\lambda_1^- \alpha + \lambda_2^- \beta
)V_- |\Psi_{b}(0)\rangle], \label{ac}
\end{eqnarray}
where $|\Psi_{b}(0)\rangle$ denotes the bath state. Then we
perform a measurement that discriminates between the encoding
space Span$\{|01\rangle,|10\rangle \}$ and Span
$\{|00\rangle,|11\rangle\}$. This measurement can be
implemented by XORing each qubit to an
ancilla qubit consecutively  \cite{pres}.
By frequently (i.e., $N$ is made
large) repeating time evolution by the Hamiltonian
in Eq. (\ref{ac}) and the consecutive measurements, we can
make effects of the terms involving $|00\rangle$ and
$|11\rangle$ in Eq. (\ref{ac}) negligible (QZE).
Then, after some simple calculations, we obtain
\begin{eqnarray}
|\Psi(T_0)\rangle &\approx& (\alpha |01\rangle + \beta|10\rangle)
\otimes |\Psi_{b}\rangle 
\nonumber\\
&&+ O(T_0)(-\alpha |01\rangle+\beta
|10\rangle) \otimes |\Psi_{b}^{\prime}\rangle, \label{af}
\end{eqnarray}
where $ |\Psi_{b}\rangle$ and $|\Psi_{b}^{\prime}\rangle$
are some arbitrary bath states which are not necessarily
orthogonal to each other. We can see that overall time
evolution generates only phase errors in the encoded bit. In
other words, the QZE prevents leakage of the states out of the
encoding space while does not prevent time evolution within
the encoding space. However, the QZE can be practicable for
only fairly stable quantum states. Typical systems are well
described by the model assumed here and thus the encoded
qubit is fairly stable. Therefore, in this case the
QZE can be a
suitable choice for protecting the encoded qubit. When
$\Delta
\epsilon$ and $\Delta \lambda^z V_z$ are negligible, the
second term of the right hand side of Eq. (\ref{af}) is
negligible and thus we need no more MRD. The two-qubit DFS in
Eq. (\ref{ad}) plus the
QZE is sufficient for preservation of one
qubit. When they are not, we should reduce the effect of the
term. This can be done in two ways, as noted
in the Introduction. First, we concatenate the DFS in
Eq. (\ref{ad}) with a three-qubit QECC that corrects only the
phase
errors (Eq. (15) of \cite{knil}). In this case, six qubits are
needed to encode one qubit in the proposed scheme. Secondly,
we concatenate the DFS in Eq.(\ref{ad}) with
a two-qubit quantum code that detects only phase errors.
That is,
\begin{eqnarray}
|0_{enc}\rangle&=& (|0\rangle+|1\rangle)
(|0\rangle+|1\rangle), \nonumber\\
|1_{enc}\rangle&=& (|0\rangle-|1\rangle)
(|0\rangle-|1\rangle),
\end{eqnarray}
where $|0\rangle$ and $|1\rangle$ denote encoded qubits
using the two-qubit DFS in Eq. (\ref{ad}) and normalization
factors are omitted. Then we preserve the
states using the QZE again: We frequently perform measurements
that tell us whether
the error has occurred or not \cite{vaid}. In
this case, four qubits are needed to encode one qubit in the
proposed scheme. The first proposed scheme (two-qubit DFS $+$
QZE) is efficient when replica asymmetry is negligible
($\Delta \lambda_z \approx 0$ and $\Delta
\epsilon \approx 0$) and other terms ($\lambda_i^{+}$ and
$\lambda_i^{-}$) are small but non-negligible. The second
([two-qubit DFS $+$ QZE]$\times $ three-qubit QECC) and third
([two-qubit DFS $+$ QZE]$\times $ [two-qubit quantum error
detecting code $+$ QZE]) proposed schemes are efficient when
replica asymmetry is also non-negligible.

In Duan and Guo's scheme \cite{dua2}, the subspace that is
orthogonal to the space to which the subspaces leak through
the 
interaction Hamiltonian is adopted as the encoding space.
Then the
QZE is used for preventing leakage of qubits out of the
encoding space. In contrast, in our scheme the qubit is
first stabilized using DFS and then leakage is prevented by
the QZE and time evolution within the encoding space is
corrected by other MRDs (QECC or quantum error detecting
code plus QZE). So the encoding space of Duan and Guo's
scheme differs from that of our scheme for a given
Hamiltonian. For example, in the case of the model of
Eq. (\ref{ab}), the encoding space of Duan and Guo's scheme
is Span$[|\bar{0}\bar{1}\rangle$, $|\bar{1}\bar{0}\rangle]$
where $|\bar{0}\rangle =(1/\sqrt{2})
(|0\rangle +|1\rangle)$ and $|\bar{1}\rangle
=(1/\sqrt{2})(|0\rangle -|1\rangle)$.
This differs from
the encoding space Span$[|01\rangle,|10\rangle ]$ of our
scheme. Duan and Guo's scheme \cite{dua2} is
more powerful than ours in that theirs is effective
for wide classes of decoherence, i.e., for independent and
even cooperative decoherence. In contrast, our scheme is
a specialized one that is efficient in the case where
phase errors are dominant but other errors are still
non-negligible.

Here we proposed three schemes that are efficient when each
pair of qubits undergoes some imperfect collective
decoherence with different baths. In the first scheme, each
pair of qubits is encoded in a DFS composed of two qubits.
Leakage out of the encoding space generated by the
imperfection is reduced by the quantum Zeno effect. In the
second scheme, phase errors in the encoded bits also
generated by imperfection of replica symmetry are reduced by
concatenation of the DFS with a three-qubit
QECC that corrects only
phase errors. In the third scheme, the same thing is
done by concatenation of the DFS with two-qubit quantum error
detecting code that detects only phase errors plus the QZE
again.

\acknowledgments
This work was supported by the Korean Ministry of Science
and Technology through the Creative Research Initiatives
Program under Contract No. 98-CR-01-01-A-20.

\end{document}